\documentclass[10pt]{IEEEtran}

\usepackage{graphicx}
\usepackage{tikz}
\usepackage{amsmath}
\usepackage{amssymb}
\usepackage{times}
\usepackage{subfigure}
\usepackage{caption} 
\usepackage{algorithm}
\usepackage{algorithmic}
\usepackage{tabularx}
\usepackage{xspace}
\usepackage{booktabs}
\usepackage{multirow}
\let\labelindent\relax
\usepackage{enumitem}
\setlist{nosep}
\usepackage{orcidlink}

\hypersetup{
  colorlinks=false,
  linkbordercolor=white,
 urlbordercolor=white,
pdfborder={0 0 0}
}

\newcommand{\includeimg}[1]{\includegraphics[width=1.0\linewidth]{#1}}
\newcommand{\textred}[1]{\textcolor{black}{#1}}

\begin{document}

\title{Activate Me!: Designing Efficient Activation Functions for Privacy-Preserving Machine Learning with Fully Homomorphic Encryption}

\author{
Nges Brian Njungle \orcidlink{0009-0006-3393-6851} and Michel A. Kinsy  \orcidlink{0000-0002-1432-6939} \\
Secure, Trusted, and Assured Microelectronics (STAM) Center\\
Ira A. Fulton Schools of Engineering\\
Arizona State University, Tempe, AZ 85281, USA\\
Emails: nnjungle@asu.edu, mkinsy@asu.edu
}

\maketitle 

\begin{abstract}
The rapid integration of machine learning into applications in sensitive domains like healthcare and defense raises serious privacy and security concerns. These applications require strong privacy protections, as they rely on large amounts of sensitive data for both training and inference.
Fully Homomorphic Encryption (FHE) offers a promising solution by allowing computations on encrypted data, thereby preserving confidentiality throughout the entire machine‑learning pipeline.
However, FHE only supports linear operations natively. This poses challenges for implementing non-linear activation functions, which are crucial for modern machine learning applications, under FHE constraints.

In this study, we design, implement, and evaluate activation functions optimized for FHE-based machine learning applications. We focus on two widely used functions: the Square function and the Rectified Linear Unit (ReLU). Our experiments utilize the LeNet-5 and ResNet-20 architectures implemented with the CKKS scheme from the OpenFHE library. 

For ReLU, we compare two approaches. First, we explore the popular low‑degree polynomial approximation approach. Second, we introduce a novel scheme‑switching technique that also securely evaluates ReLU under FHE constraints.
Our results show that the square function is highly effective in shallow models like LeNet-5, achieving 99.4\% accuracy with an inference time of 128 seconds per image. In deeper networks like ResNet-20, ReLU is more appropriate. The FHE-based ResNet-20 model implemented with ReLU polynomial approximation resulted in an accuracy of 83.8\% and an inference time of 1,145 seconds per image. Further, our proposed scheme-switching algorithm achieved a higher accuracy of 89.8\%, with an increased inference time of 1,697 seconds per image on a ResNet-20 model as well.
These results highlight the key trade-off to be considered when selecting activation functions for FHE-based machine learning applications.  The activation functions that reduce computation time tend to significantly lower accuracy, while those that preserve accuracy incur higher computational costs. 
\end{abstract}
\section{Introduction}
\label{sec:intro}

Machine learning (ML) has emerged as a cornerstone of Artificial Intelligence (AI) and is often regarded a key driver for the Fourth Industrial Revolution \cite{magd2022artificial}, \cite{virmani2024machine}. Over the past decade, it has experienced unprecedented growth, becoming an essential asset to technology across various sectors, including healthcare, finance, and defense.  The ability of ML to derive insights from vast datasets has proven invaluable in critical systems such as medical diagnostics, predictive healthcare, and military decision making \cite{virmani2024machine}. 
In defense, for example, ML is used in high-stakes applications such as autonomous weapon systems, threat detection, cybersecurity, and intelligence analysis \cite{alcantara2023evaluating}. These capabilities enable faster, data-driven decision making while significantly enhancing operational efficiency and mission effectiveness. 
However, the successful deployment of ML in these critical applications is highly dependent on access to large volumes of sensitive data for both training and inference. 
This dependency also presents serious privacy and security challenges. The privacy and security of data are particularly acute in military applications, since the data involved are often classified or highly confidential. Unauthorized access, leakage, or misuse of these data could compromise national security, operational plans, or strategic assets. As a result, ensuring the security and privacy of data in ML workflows in military systems is not only a technical necessity but a matter of national security \cite{tariq2020review}.

Traditional cryptographic methods, although effective at securing data during storage or transmission, fall short when it comes to protecting data during processing. In most cases, data must be decrypted before it can be processed, potentially exposing it to unauthorized access \cite{salavi2019survey}. Fully Homomorphic Encryption (FHE) has emerged as a promising solution to this drawback by enabling computations directly on encrypted data \cite{holygrail}. It ensures that privacy and security are maintained throughout the entire life cycle of data, making it particularly valuable for privacy-preserving machine learning (PPML) applications, where data security and privacy are paramount.
Despite its potential, the application of FHE in ML models presents two key challenges: a very high computational cost of evaluating models and the inability to efficiently evaluate non-linear functions \cite{pulido2021privacy}. 
Artificial neural networks are the dominant sub-field of ML today. These networks depend on non-linear activation functions to introduce complexity and enable effective learning. However, the FHE schemes of today typically support only linear operations. This limitation severely restricts the size and practical application of artificial neural networks under FHE constraints \cite{njungle2025guardianml}.

Prior research on the intersection of ML and FHE has centered on enhancing computational efficiency through various optimizations in the design of FHE-based models. In contrast, this work is focused on the design and implementation of activation functions tailored for FHE-based ML models. We examine two commonly used activation functions in FHE-based neural networks architectures: the square function and the Rectified Linear Unit (ReLU).
Early FHE-compatible models predominantly employed simple polynomial activations, with the square function being the most widely used. However, recent advancements in the field have shifted FHE-based models toward the adoption of polynomial approximations of the ReLU activation function \cite{banerjee2019empirical}. 
While the approximation of ReLU approach resulted in performance improvements, there have been little systematic investigation regarding the contexts and model architectures where different activation functions perform best under FHE constraints.
This work bridges that gap by implementing FHE-based neural networks using various activation functions and rigorously evaluating their effectiveness.

We evaluate the performance of these different activation functions using the LeNet-5 and ResNet-20 architectures. These architectures represent shallow and deep artificial neural network models, respectively. For the non-linear ReLU activation function, we implement the widely adopted polynomial approximation technique. We also introduced a novel algorithm that leverages FHE scheme-switching capabilities to enable a high precision evaluation of ReLU under FHE constraints. 
We compare the approaches based on inference latency and accuracy, emphasizing the trade-offs involved in designing activation functions for FHE-based ML models.
Our results indicate that the square function is highly effective for shallow networks like LeNet-5, while ReLU proves more suitable for deeper architectures like ResNet-20. 
The polynomial approximation-based approach delivered a relatively low inference latency but introduced a higher accuracy degradation. 
In contrast, our scheme-switching method achieved a higher model accuracy with minimal  degradation, albeit at the cost of increased computational overhead.
These findings provide important insights into the design of FHE-compatible activation functions and contribute to the development of more practical yet efficient PPML models and solutions.
Our main contributions in this work are as follows:
\begin{itemize}
    \item Design and Implementation of Activation Functions for FHE-based ML models namely, the square and ReLU activation functions.
    \item Introduction of a novel algorithm using the scheme-switching capabilities of FHE to evaluate the ReLU activation function in the encrypted domain.
    \item We conduct a comprehensive evaluation of the square function, the polynomial approximation of ReLU, and the proposed scheme-switching ReLU method using LeNet-5 and ResNet-20 models under FHE constraints. We highlight the trade-offs of these approaches in terms of model accuracy and inference latency.

\end{itemize}

\section{Related Works}
\label{sec:related_works}

Artificial neural networks have become the dominant approach in modern machine learning applications. They are generally composed of interconnected linear layers and activation functions. 
Activation functions play a crucial role in these models as they enable them to capture complex, non-linear patterns in data \cite{sharma2017activation}. 
In recent years, artificial neural network models incorporating FHE for enhanced security and privacy have made significant advancements. 
However, research has predominantly focused on enhancing the efficiency of encrypted inference. While notable advancements have been achieved in this direction, training artificial neural networks under FHE constraints remains an open challenge. 
This is because of the substantial computational overhead and inherent accuracy loss of models when trained in the encrypted domain. 
The high computational cost and accuracy loss associated with FHE-based training arises from the complexity of operations like backpropagation, which must be executed on large amounts of encrypted data while maintaining high precision. Thus, making the process a very expensive and challenging computational task \cite{Podschwadt}. 

PPML applications utilizing FHE are generally categorized into two generations. 
The first-generation approaches are characterized by limited model depth and high computational overhead.  The  second-generation methods introduced advanced cryptographic optimizations including approximation of ReLU and Bootstrapping. These advancements significantly improved efficiency and enabling more practical applications of FHE in machine learning applications. 
There are other existing FHE-based ML applications which leverage other privacy preserving techniques like multi-party computation \cite{catalano2005multiparty} to facilitate the evaluation of non-linear activation functions with examples being Gazelle \cite{gazelle}, MiniONN \cite{liu2017oblivious}, and XONN \cite{riazi2019xonn}. However, these works are generally considered as hybrid due to their integration of multiple  techniques to provide privacy and security thus we do not employ this approach in this work. In this work, we focus on strict FHE-based ML applications specifically artificial neural networks.

The first generation of FHE-based neural networks began with CryptoNets from Microsoft Research in 2016. The framework achieved a  99\% accuracy on the MNIST dataset using a neural network custom architecture designed specifically for this task \cite{gilad2016cryptonets}. CryptoNets main adjustment to neural network architectures was the use of the square function as their activation function which removed the complexity involved with non-linear activation functions. In 2017, Ehsan et al. proposed CryptoDL, which shitted from the simple square to a more complex low degree polynomial as the activation function \cite{hesamifard2017cryptodl}. Another notable contribution came from Al Badawi et al. who accelerated HE inference using GPUs with a custom AlexNet-like architecture \cite{al2020towards} still utilizing the square activation function. Other works in this generation such as; E2DM \cite{e2dm}, and DiNN \cite{DiNN} have also adopted the approach of using simple low-degree polynomial functions as activation functions. 

The second generation of FHE-based neural networks introduced more sophisticated architectures with significant performance improvements. These works employed sophisticated techniques like bootstrapping which enables computations of arbitrary depth networks. They also introduced the evaluation of non-linear activation functions using polynomial approximation methods to enable privacy-preserving computation in complex and deep neural networks under FHE constraints.
During training, non-linear functions are utilized (specifically the ReLU activation function). For encrypted inference, the non-linear functions are replaced with their polynomial approximations. 
A primary focus of various works in this generation has been optimizing resource usage and performance in HE-based neural networks. 
Works in this generation generally adopt the CKKS or TFHE homomorphic encryption schemes. The TFHE scheme is known for its fast and efficient bootstrapping after every multiplication called programmable bootstrapping \cite{tfhe}. 
This approach enables the evaluation of binary neural networks of arbitrary depth circuits while maintaining small ciphertext sizes with minimal precision loss. 
The most recent and most efficient of these works is the work of  Benamira et al. \cite{badawi2023}, which proposed a TFHE-based architecture that achieved 74.1\% accuracy on the CIFAR-10 dataset using a custom TFHE-CNN friendly architecture. 
The main limitation of these works stems from the fact that TFHE is inherently slow in large-scale data processing compared to other FHE schemes, like CKKS and BGV. This is due to its lack of support for parallel data processing. Also, the TFHE scheme operates on bit-wise data, which contrasts the floating-point arithmetic generally used in modern machine learning models applications. These limitations make TFHE not very tractable for processing complex or deep neural network architectures \cite{narumanchi2017performance}.
On the other hand, the CKKS scheme has emerged as the most compatible FHE scheme for ML works given that it is the only mainstream FHE scheme that supports floating-point arithmetic and parallel data processing  \cite{ckks} \cite{lee2022privacy}. These capabilities significantly accelerate computations while providing a possible direct mapping of functionalities with unencrypted ML works. In 2022, Lee et al. proposed an FHE inference system that achieved 91.31\% accuracy on the CIFAR-10 dataset in 2,271 seconds \cite{lee2022privacy} using the polynomial approximation of the ReLU activation function. Building on this, Kim et al. introduced a more efficient approach, achieving 92.04\% accuracy on CIFAR-10 with a single image inference time of 255 seconds \cite{kim2023optimized}. Most recently, Rovida et al. proposed an optimized implementation of Kim et al.’s work, achieving 91.53\% accuracy on the CIFAR-10 dataset in 260 seconds while using only 15.1 GB of memory \cite{rovida_cnn}. 
All works mentioned have used the polynomial approximation of the ReLU activation function to evaluate ResNet-20 models.

Unlike most FHE-based ML works today, which primarily focus on optimizing FHE inference while applying the approximation of ReLU or using low-degree polynomials like the square function, our work takes a different approach. We focus on evaluating different activation functions suitable for PPML solutions under FHE constraints. By analyzing different methods, we identify where the different activation functions are most effective in constructing FHE-based models. Through a comprehensive analysis of results, this work provides insights that will assist in selecting activation functions that best fit the privacy preserving needs of different applications.
\section{Background}
\label{sec:background}

\subsection{Fully Homomorphic Encryption}
Homomorphic Encryption (HE) is a cryptographic paradigm that enables direct computations on encrypted data without requiring decryption. This concept was originally introduced in 1978 under the term \textit{Privacy Homomorphism} \cite{he_init}, shortly after the introduction of the RSA public key encryption scheme \cite{holygrail}. The introduction of HE marked a significant advancement in secure computation, allowing encrypted data to be manipulated while preserving confidentiality.
A major breakthrough in this sub field of cryptography occurred in 2009 when Craig Gentry introduced the first Fully Homomorphic Encryption (FHE) scheme \cite{gentry} in his PhD thesis.  FHE enables an arbitrary number of computations on encrypted data, overcoming the limitations from earlier HE schemes that only supported a restricted set of operations and only very limited depth circuits. 
Gentry's work laid the foundation for practical FHE by introducing two fundamental techniques: \textit{bootstrapping} and \textit{squashing}. These innovations provided a structured approach to extending somewhat homomorphic encryption schemes to fully homomorphic ones.
Gentry's scheme was based on ideal lattices, that inherently support homomorphic addition and multiplication for circuits of very limited depth. 
A critical challenge and limitation of lattice based HE schemes is noise accumulation within ciphertexts. As computations proceed, the noise level increases, and if it surpasses a certain threshold, the ciphertext becomes undecipherable thus rendering the encrypted message to be irretrievable. To address this issue, Gentry used an expensive  computational operation called bootstrapping. He defined bootstrapping as a mechanism that periodically reduces noise within a ciphertext, thereby enabling indefinite computations on encrypted data. This approach established a foundation for subsequent improvements in FHE schemes, making them increasingly efficient and practical for real-world applications. 

\begin{figure}[http]
    \begin{center}
    \includeimg{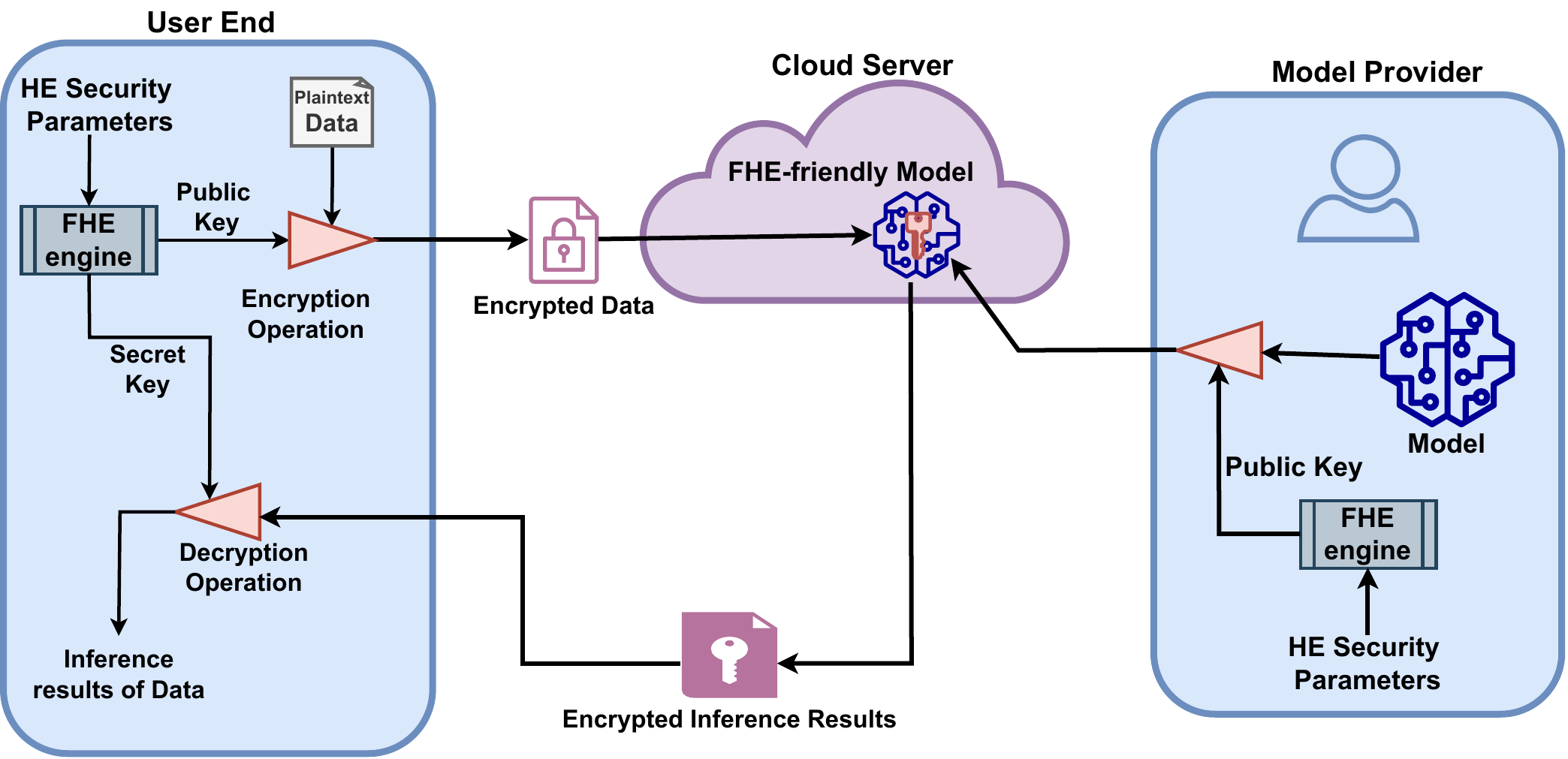}
    \captionsetup{justification=centering}
    \caption{Privacy-preserving machine learning inference in an outsourced cloud scenario showing a user inferring encrypted data over an encrypted model.}
    \label{fig:fheworks}
    \end{center}
\end{figure}

While other mainstream FHE schemes such as Brakerski-Gentry-Vaikuntanathan (BGV) \cite{bgv}, Brakerski/Fan-Vercauteren (BFV) \cite{bfv}, and Fast Fully Homomorphic Encryption over the Torus (TFHE) \cite{tfhe} exist, the Cheon-Kim-Kim-Song (CKKS) scheme \cite{ckks} is the most suitable for machine learning applications. It is the only mainstream FHE scheme that supports both approximate arithmetic and parallel data processing, therefore being the only scheme that aligns perfectly well with the needs of modern machine learning workflows and workloads. 
Figure \ref{fig:fheworks} illustrates a  privacy preserving  outsourced ML inference scenario utilizing HE. In this setup, a user is seeking to perform inference on an encrypted model stored in the cloud. First, they encrypt their input data using their public key and transmit it to the cloud. The cloud server processes the encrypted data using the encrypted model without ever decrypting it. Once the processing is completed, it returns the encrypted inference result to the user. Upon receiving the encrypted result, the user then decrypts it using their private key to gain access to the model's inference output. It is important to note that, at no point in this process did the cloud provider or anyone else gain access to the input data or inference result except for the user. 
Also, the cloud provider cannot extract any information from the encrypted model, since they do not have access to the private key.
Furthermore, even in the event of a cloud breach, the user data and model stay protected, since all information is maintained in its encrypted form.
This mechanism upholds confidentiality of both the user's data and the model, making HE a robust solution for secure cloud-based machine learning inference.

\subsection{Cheon-Kim-Kim-Song (CKKS) Scheme}

In 2016, Cheon, Kim, Kim, and Song introduced the CKKS scheme, a HE scheme specifically designed for approximate arithmetic \cite{ckks}. Its security is based on the Ring Learning with Errors (RLWE) problem, a structured extension of the Learning with Errors (LWE) problem proposed by Regev in 2005 \cite{lwepropsed} and later refined for polynomial rings by Lyubashevsky et al. \cite{ringlwe}.  
The LWE problem involves recovering a hidden vector from noisy linear equations generated using random samples. Formally, given a matrix \( A \in \mathbb{Z}_q^{m \times n} \) and a vector \( b \in \mathbb{Z}_q^m \), the goal is to find an unknown vector \( s \in \mathbb{Z}_q^n \) satisfying:  
\begin{equation}
    A s + e = b \; \mathrm{mod} \; q,
\end{equation}  
where \( e \) is a noise vector sampled from a predefined error distribution, and \( q \) is a large prime modulus.  

Expanding on this, the RLWE problem generalizes LWE to polynomial rings, leveraging their algebraic properties for computational efficiency. A key advantage of CKKS is its ability to encode multiple plaintext values into a single polynomial through a technique known as ``slot packing". This feature enables efficient parallel computations using the Single Instruction Multiple Data (SIMD) paradigm, significantly accelerating homomorphic operations through  batch processing of data.  
The CKKS scheme defines a range of homomorphic operations, including addition, subtraction, multiplication, rotation, and bootstrapping. These operations enable users to build a wide range of applications that support encrypted data processing with CKKS making it well-suited for PPML and other critical applications that require privacy computations on real-valued data.
\subsection{Machine Learning}

Machine learning (ML) is a sub-field of artificial intelligence that empowers computers to recognize patterns within data and generate predictions or decisions without requiring explicit programming \cite{MohammadAlRubaie-2019}. 
The field of ML is broadly categorized into three paradigms: supervised learning, unsupervised learning, and reinforcement learning. Supervised learning involves training models using labeled datasets, enabling them to make precise predictions based on known input-output pairs \cite{ZoubinGhahramani-2004}. In contrast, unsupervised learning identifies hidden structures and patterns within unlabeled data, uncovering insights without predefined outcomes \cite{TammyJiang-2020}. Reinforcement learning is distinct from these approaches. It employs a trial-and-error mechanism where models interact with an environment, then develop optimal strategies through rewards and penalties \cite{KeerthanaSivamayil-2023}.

A significant advancement within the field of ML is the area of deep learning. Deep learning utilizes artificial neural networks (mainly convolution neural networks) with multiple layers to process complex and high-dimensional data representations  \cite{AlexKrizhevsky-2012}. These networks typically comprise of linear layers and activation functions. The linear layers generally include the convolution layers, the pooling layers, and the fully connected layers. These linear layers have been widely studied under FHE constraints as they generally account for the bulk of the computational cost of FHE-based models. For this reason, we will not discuss them in this work. Instead, we will focus on activation functions since they have not been well studied under FHE constraints. Activation functions are particularly interesting as they introduce nonlinearity in learning, allowing ML model to capture intricate relationships within data. Commonly used activation functions in modern ML applications include the Rectified Linear Unit (ReLU), Sigmoid, and Hyperbolic Tangent (Tanh), each contributing uniquely to model efficiency and performance. Deep learning techniques can be applied across all three ML paradigms, enhancing supervised learning with highly accurate models, enabling advanced clustering in unsupervised learning, and improving decision-making capabilities in reinforcement learning.

\section{FHE-based Activation Functions}
\label{sec:activationfunctions}

Activation functions play a fundamental role in ML models. Particularly in artificial neural networks, they introduce non-linearity and enabling complex pattern learning \cite{DUBEY202292}. This non-linearity allows networks to learn complex patterns and relationships in data that linear models cannot capture. By determining how neurons process and transmit signals through layers, activation functions fundamentally shape a network’s ability to learn and generalize from data.
Traditional activation functions like the sigmoid and hyperbolic tangent (tanh) were widely used in early neural networks due to their smooth, differentiable properties. However, they often suffered from vanishing gradient issues, which made training deep networks difficult \cite{roodschild2020new}, \cite{hu2018overcoming}. In response, more recent advancements have favored the Rectified Linear Unit (ReLU) and its variants because they provide faster convergence, reduce the gradient-related problems, and encourage sparse activations~\cite{gupta2020effect}, \cite{he2018relu}.

While activation functions are relatively straightforward to implement in the plaintext domain, they present significant challenges in PPML settings, particularly when using FHE. For instance, ReLU, is non-polynomial and non-smooth, making it incompatible with the algebraic operations supported by most FHE schemes. To address this issue, researchers have proposed FHE-based alternatives based on low-degree polynomial functions including  simple square functions and polynomial approximations of ReLU. 
In this section, we examined the design and implementation of these activation functions in the FHE context, highlighting their algorithms, benefits, and the trade-offs they entail.

\subsection{Square Activation function}

Popular activation functions used in ML, such as the ReLU and sigmoid functions, involve non-polynomial operations that are difficult to compute under FHE constraints. In contrast, the square function is inherently polynomial, making it well-suited for homomorphic evaluation. Its algebraic simplicity aligns naturally with the linear structures supported by most FHE schemes, making it both easy to understand and straightforward to implement.
Due to these advantages, the square activation function gained significant attention in the context of FHE-based PPML solutions. As a low-degree polynomial, it was the first activation function adopted for encrypted neural networks computation. Among other low-degree polynomials, it is valued for its simplicity, computational efficiency, and compatibility with homomorphic operations.
Mathematically, the square activation function is defined in Equation~\ref{eq:square}. 
\vspace{-0.05in}
\begin{align}
    f(x) = x^2 \label{eq:square}
\end{align}
where $x$ is the input to the function $f$.
\vspace{0.1in}

The implementation of the square activation function using the CKKS scheme is outlined in Algorithm~\ref{algo:square}.
\begin{algorithm}
\caption{The Square Function Algorithm in FHE}
\label{algo:square}
\begin{algorithmic}[1]
\STATE \textbf{Input:} $x$ (input ciphertext)
\STATE \textbf{Output:} $y$

\STATE $y \gets \textit{CKKSmultiply}(x, x)$

\RETURN $y$

\end{algorithmic}
\end{algorithm}

The square activation function is smooth and differentiable everywhere, which facilitates stable gradient propagation during training. However, its derivative, given by \( f'(x) = 2x \), is also a linear function. Unlike non-linear activation functions whose gradients exhibit non-linear behavior promoting piecewise non-linearity and sparsity, the square function's linear derivative introduces certain optimization challenges. In particular, the lack of saturation regions and the absence of a thresholding mechanism reduce the network's capacity to capture complex hierarchical features effectively.
Another key limitation of the square function is its inherent non-negativity. By squaring all inputs, it eliminates sign information and produces only non-negative outputs without any accumulations. This behavior can lead to redundancy in activation patterns and increase the correlation among neuron outputs, which may negatively affect the model's ability to generalize.
Moreover, the unbounded nature of the derivative \( 2x \) for large input magnitudes can cause gradient instability during training. This is especially problematic in deep architectures, where the accumulation of large gradients can result in exploding gradient phenomena, ultimately destabilizing the learning process.

These challenges make it difficult to train deep neural networks effectively using the square activation function. Modern architectures like ResNet-20, rely on residual connections that perform best when activation functions maintain controlled gradient norms. The unbounded growth of the square function's derivative disrupts this balance, leading to unstable weight updates.
As a result, while the square function is computationally efficient and works well in shallow FHE-based models, it is not suitable for deeper architectures. For instance, training the ResNet-20 model with the square activation function was proved infeasible due to instability during backpropagation. The exploding gradients prevented effective learning and completely hindered the network's ability to generalize.

\subsection{ReLU Activation Function}

Rectified Linear Unit (ReLU) is the most widely adopted activation function in ML today due to its simplicity and effectiveness \cite{he2018relu}. ReLU applies a threshold operation where all negative values are set to $0$. This non-linearity accelerates convergence and helps reduce the vanishing gradient problem in training. Equation \ref{eq:relu} shows the mathematical representation of the ReLU activation function. 
\begin{align}
\label{eq:relu}
    f(x) = max(0, x) 
\end{align}

Unlike traditional activation functions such as sigmoid and tanh, that suffer from vanishing gradient problems, ReLU preserves gradient flow for positive inputs, allowing for faster convergence and improved performance in deep architectures. By setting all negative inputs to zero, ReLU introduces sparsity in activations, reducing computational complexity and enhancing model generalization by mitigating unnecessary neuron activations. This sparsity also contributes to an implicit form of regularization, preventing over-fitting in large networks. 

\subsubsection{Polynomial Approximation of ReLU}
In FHE-based ML models, the implementation of ReLU poses challenges due to its non-polynomial nature, making direct computation on encrypted data inefficient especially in the CKKS scheme.  Previous FHE-based neural networks implementations have focused on Polynomial approximation of the ReLU activation function mainly using the Chebyshev's Polynomial Approximation.

Chebyshev polynomials form a family of orthogonal polynomials defined on the interval \([-1, 1]\). They are highly effective in approximating complex, non-linear functions with both efficiency and precision \cite{gil2012non}. These polynomials are denoted by \(T_n(x)\) and are defined recursively as follows:
\begin{align}
    T_0(x) &= 1, \\
    T_1(x) &= x, \\
    T_n(x) &= 2x T_{n-1}(x) - T_{n-2}(x), \quad \text{for } n \geq 2.
\end{align}

A distinctive feature of Chebyshev polynomials is the distribution of their roots. A polynomial of degree \(n\) has exactly \(n\) roots, all of which are located within the interval \([-1, 1]\). These roots are expressed in the following form:
\begin{align}
    x_k = \cos\left(\frac{k\pi}{n}\right), \quad \text{for } 0 \leq k < n. \label{eq:cheroots}
\end{align}

The  Chebyshev polynomials offer an efficient way to express non-linear functions as a sum of polynomials. This approach proves especially useful in FHE applications as polynomials  seamlessly integrate with the underlining data representations. By employing Chebyshev polynomial approximations, the ReLU function can be approximated securely within the encrypted domain.

To correctly apply Chebyshev approximations, it is important to normalize the input data to lie within the interval \([-1, 1]\). We achieved this by introducing a scaling factor \(\beta\) that adjusts the input vector accordingly. The input vector is scaled by multiplying it with \(\frac{1}{\beta}\), as shown in Equation \ref{eq:scale_relu}:
\begin{align}
\label{eq:scale_relu}
    x_{\text{s}} = \frac{1}{\beta} \cdot x
\end{align}

The parameter \(\beta\) is selected through an analysis of values across different datasets and networks. This process ensures that the data is accurately normalized while maintaining the representational integrity of the ReLU activation function under FHE constraints. When \(\beta > 1\), a scaling mask is applied to the input vector \(\mathbf{x}\) to ensure proper scaling of data before processing. Once scaled, the modified input is passed to the Chebyshev approximation function. The results are then scaled back by multiplying them by \(\beta\), as shown in Algorithm~\ref{alg:secure_relu}.

\begin{algorithm}
\caption{Secure ReLU Using Homomorphic Encryption}
\label{alg:secure_relu}
\begin{algorithmic}[1]
\STATE \textbf{Input:} $x$, $\beta$
\STATE \textbf{Output:} $y$

\STATE \textbf{Initialization:}  
 $D $, $v \gets x$

\IF{$\beta > 1$}
    \STATE $v \gets \text{CKKSMultiply}(x, \beta)$
\ENDIF

\STATE Define $f(z): f(z) \gets 0 \text{ if } z < 0; \; f(z) \gets \beta \cdot z \text{ otherwise}$

\STATE $y \gets \text{ChebyshevFunction}(f(z), v, D)$

\STATE \textbf{return} $y$
\end{algorithmic}
\end{algorithm}

In Algorithm \ref{alg:secure_relu}, \(D\) is the degree of the Chebyshev polynomial used for approximating the function.
The value of $D$ determines the accuracy of the Chebyshev Approximation of the ReLU function. The higher the degree, the better the approximation but the higher the computational cost. 
To determine the right value of D to use, we conducted an experiment with varying values for the polynomial degree between $10$ and $100$. We studied how well the function approximated the results of the ReLU function. We determined that when $D$ is equal $50$, we obtain a good balance between accuracy and the resources used for the polynomial approximation of the ReLU activation function. This value of $D$ consumes a noise budget equivalent to that of eight multiplications thus efficient for use in applications with FHE parameters that can lead to a multiplication depth after bootstrapping to still be greater than eight.

\subsection{Secure ReLU Evaluation via Scheme Switching}

FHE schemes exhibit varying degrees of efficiency and computational capabilities, making them suitable for different types of encrypted computations and operations. The CKKS scheme, in particular, is well-suited for approximate arithmetic and supports efficient parallel computation. These properties make CKKS the optimal choice for evaluating linear operations in FHE-based ML applications. However, CKKS is inherently limited in its ability to perform non-linear operations, such as the comparison operation required for thresholding in the ReLU activation function.

In contrast, the FHEW~\cite{fhew} and TFHE schemes \cite{tfhe} are specifically designed for efficient evaluation of boolean gates. They enable precise computation of non-linear functions such as comparisons, sign determination, and the floor function~\cite{tfhe}. These functions are expressed as boolean circuits which are then constructed from basic logic gates such as AND, OR, NOR, and XOR defined in these schemes. This gate-based approach allows for accurate discrete computation of functions under FHE constraints.
While FHEW and TFHE offer high accuracy in such boolean computations, they do not support SIMD-style parallelism. Consequently, they introduce significant computational overhead when applied to large-scale data processing tasks like those commonly encountered in ML applications.

To leverage the strengths of both encryption schemes, we adopt a scheme-switching approach that integrates CKKS with TFHE. Specifically, CKKS is used to efficiently evaluate all linear layers in the model, while TFHE is employed in the evaluation of non-linear activation functions. During inference, ciphertexts are switched back and forth between the two schemes depending on the type of layer and operation been evaluated. This hybrid strategy enables efficient computation while maintaining compatibility with the constraints of FHE and capitalizing on the complementary strengths of both schemes.
The concept of scheme switching between different homomorphic encryption schemes was first introduced by Christina et al. in 2018 under the name CHIMERA~\cite{chimera}. This technique is feasible between CKKS and TFHE because both schemes are rooted in the LWE problem. As a result, ciphertexts in either scheme share the same foundational algebraic structures, making translation between schemes mathematically viable. In essence, scheme switching involves transforming the underlying representation of ciphertexts from one scheme to another while preserving the encrypted data and its semantic integrity.

In this method of evaluating the ReLU activation function, we first determine the sign of each encrypted value within a CKKS SIMD ciphertext by switching to the TFHE scheme, which supports precise comparison operations. Once the signs are computed in TFHE, we switch back to CKKS and use the resulting ciphertext to construct a binary mask. This mask is then homomorphically multiplied with the original CKKS ciphertext, effectively zeroing out the negative values and thereby implementing a more precise, secure, and privacy-preserving ReLU activation function. This algorithm consumes just a single multiplication depth in the noise budget of ciphertext. The complete procedure is outlined in Algorithm~\ref{algo:schemeswitch_relu}.

This proposed scheme-switching technique effectively integrates the strengths of CKKS and TFHE, enabling efficient and secure ReLU activation in encrypted neural networks. CKKS provides high throughput in evaluating  linear layers, while TFHE assist in a precise computation of the comparison operation needed to effectively evaluate the non-linear ReLU activation  function. This hybrid approach allows for the construction of high-precision PPML models under FHE constraints.
Although the scheme-switching process of ReLU still results in some errors, the resulting errors are much smaller than those incur in the approximation when using low-degree polynomials.

\begin{algorithm}
\caption{Secure Scheme-Switching ReLU Algorithm}
\label{algo:schemeswitch_relu}
\begin{algorithmic}[1]
\STATE \textbf{Input:} $c_{\text{enc}}$ (encrypted input vector), $c_{\text{sec}}$ (second ciphertext), $\text{vector\_size}$
\STATE \textbf{Output:} $c_{\text{result}}$

\STATE Set slots of $c_{\text{enc}}$ to $\text{vector\_size}$
\STATE $tc_{\text{enc}} \gets \text{CKKSSwitchToTFHEW}(c_{\text{enc}})$
\STATE $tc_{\text{sec}} \gets \text{CKKSSwitchToTFHEW}(c_{\text{sec}})$
\STATE $c_{\text{comp}} \gets \text{TFHECompare}(tc_{\text{enc}}, tc_{\text{sec}}, \text{vector\_size})$
\STATE $\text{mask} \gets \text{GenerateCKKSPlaintextMask}(1, \text{vector\_size})$
\STATE $c_{\text{sign}} \gets \text{CKKSSubtract}(\text{mask}, c_{\text{comp}})$
\STATE $c_{\text{result}} \gets  \text{CKKSMultiply(}c_{\text{sign}}, c_{\text{enc}})$
\RETURN $c_{\text{result}}$
\end{algorithmic}
\end{algorithm}

\section{Experiment and Results}
\label{sec:experiment}

All experiments in this study were performed on a \textred{AMD Ryzen 9 5900X 12-core processor with 64GB of RAM}. We build all our models on OpenFHE v1.2.3 which is the most stable version of the library at the time of development. We chose OpenFHE because it is the most advanced FHE library available to the development community. Additionally, it is the only FHE library that supports implementations for both CKKS and FHEW/TFHE gates \cite{OpenFHE} \cite{openfhesec}. 

\subsection{Network Architectures}  

To comprehensively evaluate the performance of these different activation function implementations under FHE, we evaluate two distinct neural network architectures: LeNet-5 and ResNet-20. These models differ significantly in complexity and application domains, making them ideal candidates for assessing the trade-offs between efficiency and accuracy in privacy-centric applications.  

\begin{figure}[http]
    \centering
    \includegraphics[width=1.0\linewidth]{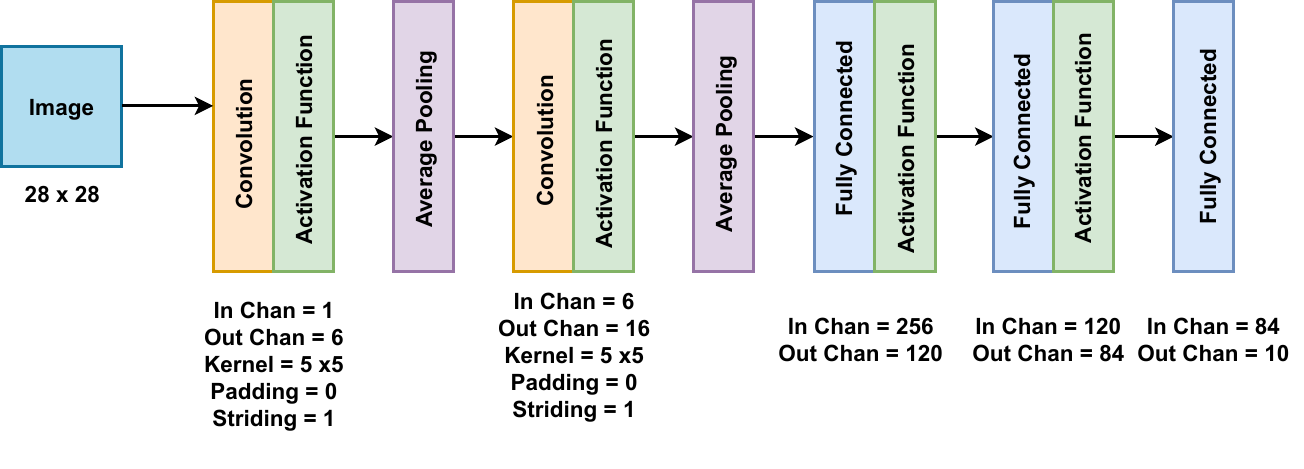}
    \captionsetup{justification=centerlast}
    \caption{The LeNet-5 Architecture used in this work: \\
    In Chan = input channels, Out Chan = output channels}
    \label{fig:lenet5}
\end{figure}

LeNet-5 is a relatively simple and shallow artificial neural network originally designed for grayscale image classification tasks such as handwritten digits recognition \cite{lenets}. Despite its simplicity, LeNet-5 remains one of the most widely studied architectures in ML literature, making it an excellent baseline for evaluating activation functions in FHE settings.  
For this study, we trained two versions of LeNet-5 on the MNIST dataset using the ReLU activation function and using the square activation function.
The training process was conducted in PyTorch, and the learned model weights were exported as CSV files which were then loaded into the encrypted models for inference. Within the encrypted domain, we implemented three different evaluation models. Our first model used the square activation function with equivalent weights exported from the model trained on the square activation function.  Our second model used the polynomial approximation of ReLU activation function while our third model used the scheme-switching evaluation of  ReLU activation function. The last two FHE-based LeNet-5 models used the weights from the model trained with the ReLU activation functions. 
Each encrypted model was evaluated using 750 images from the MNIST validation set. Figures \ref{fig:lenet5}  illustrate the detailed architectures of the LeNet-5 models used in this study including the input channels, output channels, kernel, padding, and striding configurations.

The ResNet models are some of the most widely adopted deep learning models for complex image classification tasks today \cite{he2015deepresiduallearningimage}. The core innovation in ResNets is the introduction of residual connections, which mitigate the vanishing gradient problem and enable efficient training of deeper networks. Due to its balance of depth and computational feasibility, ResNet-20 has been commonly employed in prior FHE research and serves as an ideal benchmark for our study.  

 \begin{figure}[ht]
    \centering
    \includegraphics[width=1.0\linewidth]{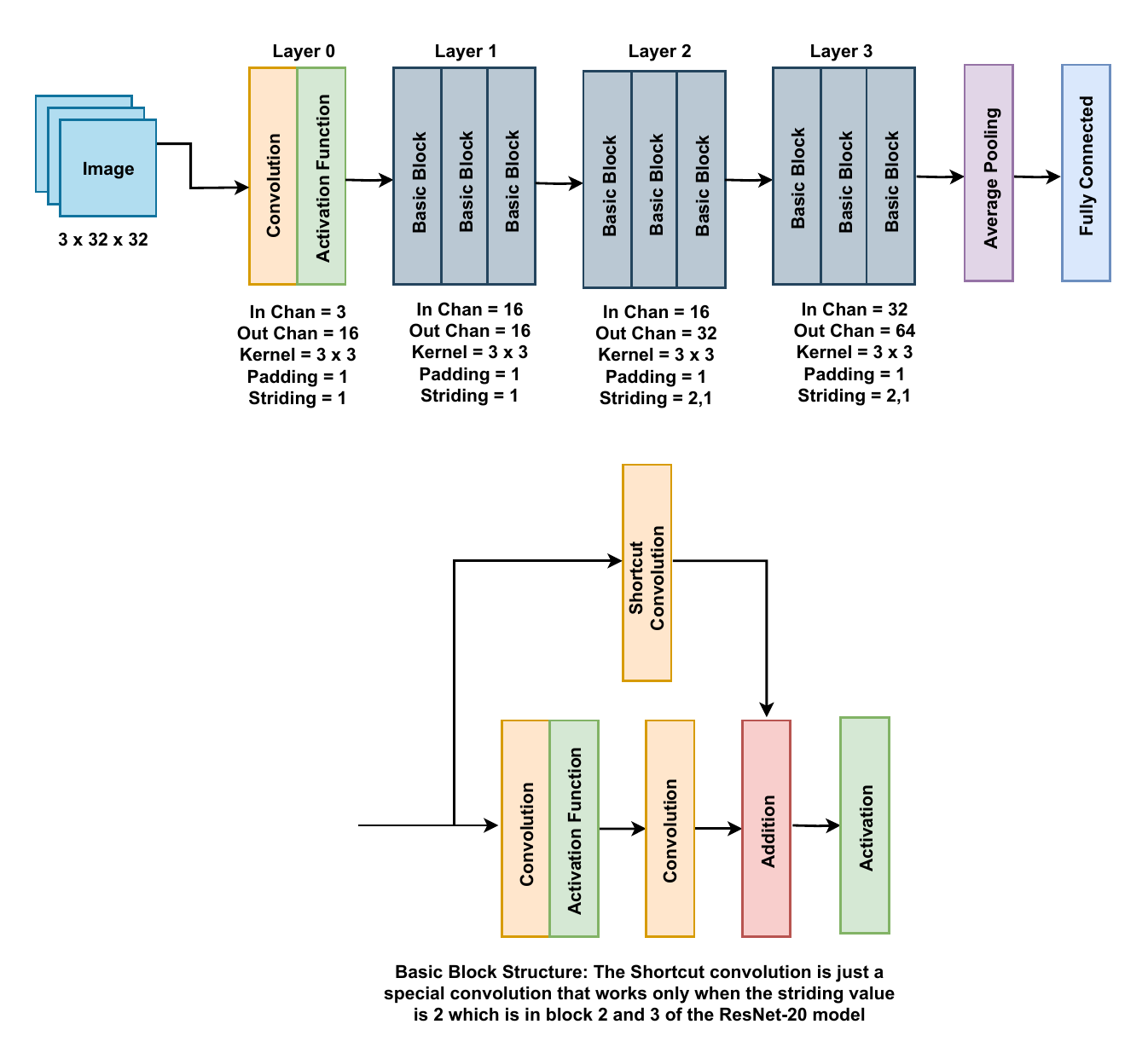}
    \caption{ResNet-20 Architecture. The Blocks show the ResNet blocks. The Block Structure shows the details and how the convolutions are connected}
    \label{fig:resnet20}
\end{figure}

For our evaluation, we trained a ResNet-20 model on the CIFAR-10 dataset using the ReLU activation function. In contrast to LeNet-5, we found that training ResNet-20 with the square activation function was infeasible due to its limited capacity to approximate non-linear behaviors in deep architectures (see Section~\ref{sec:activationfunctions}). This result highlights a fundamental limitation of low-degree polynomial activation functions in privacy-preserving deep learning, as their expressive power diminishes with increasing network depth.

After training the ResNet-20 model with ReLU activation functions, the resulting model weights were also exported as CSV files for use with the FHE-based models  to inference on encrypted data. Within the encrypted domain, we implemented two evaluation models: one utilizing a polynomial approximation of ReLU and the other employing the scheme-switching approach introduced in Algorithm~\ref{algo:schemeswitch_relu}.
We conducted encrypted inference on both models on a subset of 250 images from the CIFAR-10 validation set.
The key metrics evaluated were accuracy retention and inference latency for the two ReLU implementation approaches. To achieve fast inference for accuracy evaluation, we utilized the SOL supercomputer at Arizona State University~\cite{solsupercom}. 
Latency measurements, however, were collected directly from a local consumer-grade setup using an AMD Ryzen 9 5900X CPU.
Figures \ref{fig:resnet20} illustrate the ResNet-20 architectures used in this study including the input channels, output channels, kernel, padding and striding configurations.  Its residual connections show the additional convolutions evaluated in the down-sampling stage of ResNets. 

Furthermore, since noise accumulation is a critical challenge in FHE, we incorporated bootstrapping operations at necessary points to preserve computational integrity. Table~\ref{tab:bootstrapping} summarizes the number of bootstrapping operations required for each model to ensure accurate and efficient encrypted inference. 
The placement of bootstraps within the models were carefully determined based on the estimated noise levels at various stages of the inference process. These estimates were derived from a detailed analysis of ciphertext modulus consumption throughout the computation pipeline. By strategically applying bootstrapping only when the noise threatened to exceed the decryption threshold, we minimized overhead while maintaining correctness in the encrypted domain.

 \begin{table}[http]
    \small
    \centering
    \caption{Number of Bootstrapping Operations required in Encrypted Models}
    \label{tab:bootstrapping}
    \begin{tabularx}{\linewidth}{|X|X|}
        \hline
        \textbf{Architecture} & \textbf{Number of Bootstraps used} \\
        \hline
        LeNet-5 With Square  & 0 \\
        \hline
        LeNet-5 with Approximation & 4 \\
        \hline
        LeNet-5 with Scheme Switch & 0\\
        \hline
        ResNet-20 with Approximation   & 18 \\
        \hline
        ResNet-20 with Scheme Switching  &  8 \\
        \hline
    \end{tabularx}
\end{table}

Table~\ref{tab:fhe_params} provides an overview of the security parameters used in our encrypted inference experiments. These parameters were carefully chosen to maintain a security level around 128 bits of security with maximum throughput within the models, ensuring robustness against cryptographic attacks while optimizing computational efficiency.  

\begin{table}[ht]
    \small
    \centering
    \caption{The CKKS Parameter sets used to evaluate the LeNet-5 and ResNet-20}
    \label{tab:fhe_params}
    \begin{tabularx}{\linewidth}{|l|X|X|}
        \hline
        \textbf{Parameter} & \textbf{LeNet-5} & \textbf{ResNet-20} \\
        \hline
        Polynomial Degree & 16384 & 32768 \\
        \hline
        Number of Slots & 8192 & 16384 \\
        \hline
        Multiplications after Bootstrap & 10 & 10 \\
        \hline
    \end{tabularx}
\end{table}

\subsection{Results}

Table \ref{tab:accuracy_comparison} presents the accuracy of the evaluated neural network models under both plaintext and under FHE constraints, as well as the latency measurements in the encrypted domain. The primary objective of this evaluation was to investigate the trade-offs between computational overhead and model accuracy associated with different methods for evaluating activation functions under FHE constraints.
Our analysis compares the three activation function evaluation approaches in the encrypted domain: the low-degree polynomial square function, polynomial approximations of the ReLU function, and the scheme-switching ReLU evaluation method. The results show the inherent trade-offs between accuracy retention and computational efficiency in FHE-based PPML solutions.

For the LeNet-5 architecture, all three activation function methods achieve comparable performance, exhibiting minimal accuracy loss between plaintext and encrypted inference. Notably, the square activation function achieves the highest encrypted accuracy of $99.4\%$, with only a $0.2\%$ degradation from its plaintext counterpart. However, it offers latency performance higher than that of the ReLU polynomial approximations which significantly outperforms the scheme-switching method in terms of runtime. A key advantage of the square function lies in its seamless compatibility with the CKKS scheme, which allows it to be implemented using a single homomorphic multiplication thereby eliminating the need for bootstrapping and further improving computational efficiency in shallow architectures like LeNet-5.

The ReLU approximation method shows a modest accuracy degradation of $0.3\%$ in the encrypted LeNet-5 model while maintaining the best latency across all evaluations. Meanwhile, the scheme-switching ReLU approach also incurs a $0.3\%$ accuracy loss but introduces significant computational overhead. Specifically, it requires approximately 168 seconds to inference a single image which is about 1.7$\times$ the latency of the approximation-based method. This increased cost arises from the additional operations required to switch between CKKS and TFHE schemes, along with the inherently higher computational complexity of TFHE operations. Despite its slower performance, the scheme-switching method still offers latency that is practical for many real-world applications.

In deeper architectures like ResNet-20, the choice of activation function plays a critical role in determining the model’s performance. Residual networks, in particular, rely on non-linear activations to enable effective feature transformation across layers. Consequently, the selection of an appropriate activation function directly affects the model’s expressiveness and accuracy.
When employing polynomial approximations of the ReLU function within the CKKS framework, we observed an $8.4\%$ accuracy degradation relative to the plaintext model. This substantial drop in performance can be attributed to the limited capacity of low-degree polynomials to accurately approximate the non-linear behavior of ReLU, thereby leading to suboptimal feature transformations during inference. Nevertheless, this method remains computationally efficient, as it avoids the need for scheme-switching or bootstrapping, making it well-suited for latency-sensitive applications where minor accuracy compromises are acceptable.
In contrast, the scheme-switching ReLU evaluation method achieves significantly higher accuracy, incurring only a $2.4\%$ drop from the plaintext baseline. This demonstrates the effectiveness of leveraging TFHE’s precise operations capabilities to reconstruct the ReLU function more efficiently within the encrypted domain. However, the improved accuracy comes at the cost of increased computational complexity. The scheme-switching model required approximately $1,697$ seconds for inference, representing a $1.5\times$ increase in runtime compared to the polynomial approximation approach.

\begin{table}[http]
    \small
    \centering
    \caption{Accuracy Comparison Between Plaintext and FHE Settings as well as the latency in the FHE setting for the  LeNet-5 models and ResNet-20 models}
    \label{tab:accuracy_comparison}
    \begin{tabularx}{\linewidth}{|l|X|X|X|}
        \hline
        \textbf{Architecture} & \textbf{Plaintext (\%)} & \textbf{Encrypted  (\%)} & \textbf{Latency (s)} \\
        \hline
        LeNet-5 Square  & 99.6 & 99.4 & 128 \\
        \hline
        LeNet-5 Approximation  & 99.2 & 98.9 & 95 \\
        \hline
        LeNet-5 Scheme Switch  & 99.2 & 98.9 & 168 \\
        \hline
        ResNet-20 Approximation  & 92.2 & 83.8 & 1,145 \\
        \hline
        ResNet Scheme Switch  & 99.2 & 89.8 & 1,697 \\
        \hline
    \end{tabularx}
\end{table}

These results highlight a fundamental trade-off in PPML applications that rely on FHE. While polynomial approximations of ReLU activation functions substantially accelerate encrypted inference, they introduce notable accuracy degradation, particularly in deeper network architectures. Conversely, our introduced scheme-switching methods offer a more faithful representation of ReLU, preserving model accuracy at the cost of increased computational overhead.

With shallow neural networks like LeNet-5, polynomial activation functions,  particularly the square function, offer an optimal balance between accuracy and computational efficiency under FHE constraints. This function aligns well with the CKKS encryption scheme and avoids costly bootstrapping operations in this model, making it especially suitable for real-time applications in resource-limited settings. ReLU-based LeNet-5 models also performed well, with the approximation approach achieving the lowest latency in the FHE-based LeNet-5 models.
For deeper architectures like ResNet-20, the optimal activation function strategy depends heavily on application-specific requirements. If accuracy preservation is paramount, the scheme-switching ReLU approach provides a robust solution with minimal degradation. However, if inference latency and computational resources are the primary constraints, ReLU approximation methods offer a practical alternative, albeit with a slight compromise in accuracy.
Ultimately, these findings underscore that activation function selection in PPML is not a one-size-fits-all decision. Researchers and practitioners must carefully evaluate their deployment requirements in other to balance accuracy, latency, and model depth when selecting an appropriate activation function strategy for FHE-based neural networks.

\section{Conclusion}
\label{sec:conclusion}

In this work, we explored the impact of activation functions on the efficiency and accuracy of privacy-preserving neural networks under Fully Homomorphic Encryption (FHE) constraints. We also introduce a novel algorithm for high precision evaluation of the ReLU activation function based on scheme-switching between the TFHE and CKKS schemes. Our results, highlight the trade-offs between the square activation function, ReLU approximation, and our proposed scheme-switching ReLU activation functions in both shallow and deep neural network architectures.  
For shallow networks like LeNet-5, the square activation function proved to be the most efficient choice, offering the best accuracy with a comparable latency. Additionally, its implementation allowed us to eliminate the need for bootstrapping. This further enhance computational efficiency of models like LeNet-5 in scenarios where bootstrapping implementation is not possible. On the other hand, the challenges that come with low degree polynomial activation functions makes it difficult to adopt the square activation function in deep models. 
For deeper architectures like ResNet-20, the choice of activation function had a more pronounced effect. The polynomial approximation approach for evaluating ReLU was computationally efficient but resulted in a notable accuracy drop of $8.4\%$. In contrast, the scheme-switching method preserved accuracy with only a $2.4\%$ degradation. This lower accuracy degradation came at though the cost of increased latency.

These findings highlight a fundamental trade-off in the selection of activation functions for privacy-preserving machine learning application based on FHE. Polynomial activations like the square activation function are computationally efficient and well-suited for shallow models such as LeNet-5, whereas the scheme-switching method is better aligned with deep architectures where maintaining precision is critical. In performance sensitive applications, the ReLU approximation approach emerges as the most practical choice, underscoring the importance of fine-grained tuning to reduce accuracy degradation. In high precision settings, our proposed scheme-switching approach of evaluating the ReLU activation function is the most appropriate though it comes with a higher computational cost compared to the approximation approach. 
Thus, the choice between activation functions in FHE-based machine learning applications should be guided by the specific accuracy and efficiency requirements of the target application. Future research should prioritize optimizing scheme-switching mechanisms to reduce inference latency without compromising accuracy. Another interesting research direction is the fine-tuning of the polynomial approximation technique to reduce their accuracy degradation rates as well as hybrid applications of these techniques. These advancements will enhance the feasibility and scalability of FHE-based solutions in modern machine learning applications.

\bibliographystyle{IEEEtran}
\bibliography{paper}

\begin{thebibliography}{10}
\providecommand{\url}[1]{#1}
\csname url@rmstyle\endcsname
\providecommand{\newblock}{\relax}
\providecommand{\bibinfo}[2]{#2}
\providecommand\BIBentrySTDinterwordspacing{\spaceskip=0pt\relax}
\providecommand\BIBentryALTinterwordstretchfactor{4}
\providecommand\BIBentryALTinterwordspacing{\spaceskip=\fontdimen2\font plus
\BIBentryALTinterwordstretchfactor\fontdimen3\font minus \fontdimen4\font\relax}
\providecommand\BIBforeignlanguage[2]{{%
\expandafter\ifx\csname l@#1\endcsname\relax
\typeout{** WARNING: IEEEtran.bst: No hyphenation pattern has been}%
\typeout{** loaded for the language `#1'. Using the pattern for}%
\typeout{** the default language instead.}%
\else
\language=\csname l@#1\endcsname
\fi
#2}}

\bibitem{magd2022artificial}
H.~Magd, H.~Jonathan, S.~A. Khan, and M.~El~Geddawy, ``Artificial intelligence—the driving force of industry 4.0,'' \emph{A roadmap for enabling industry 4.0 by artificial intelligence}, pp. 1--15, 2022.

\bibitem{virmani2024machine}
D.~Virmani, M.~A.~S. Ghori, N.~Tyagi, R.~Ambilwade, P.~R. Patil, and M.~Sharma, ``Machine learning: The driving force behind intelligent systems and predictive analytics,'' in \emph{2024 International Conference on Trends in Quantum Computing and Emerging Business Technologies}.\hskip 1em plus 0.5em minus 0.4em\relax IEEE, 2024, pp. 1--6.

\bibitem{alcantara2023evaluating}
E.~J. Alc{\'a}ntara~Su{\'a}rez and V.~Monzon~Baeza, ``Evaluating the role of machine learning in defense applications and industry,'' \emph{Machine Learning and Knowledge Extraction}, vol.~5, no.~4, pp. 1557--1569, 2023.

\bibitem{tariq2020review}
M.~I. Tariq, N.~A. Memon, S.~Ahmed, S.~Tayyaba, M.~T. Mushtaq, N.~A. Mian, M.~Imran, and M.~W. Ashraf, ``A review of deep learning security and privacy defensive techniques,'' \emph{Mobile Information Systems}, vol. 2020, no.~1, p. 6535834, 2020.

\bibitem{salavi2019survey}
R.~R. Salavi, M.~M. Math, and U.~Kulkarni, ``A survey of various cryptographic techniques: From traditional cryptography to fully homomorphic encryption,'' in \emph{Innovations in Computer Science and Engineering: Proceedings of the Sixth ICICSE 2018}.\hskip 1em plus 0.5em minus 0.4em\relax Springer, 2019, pp. 295--305.

\bibitem{holygrail}
D.~Tourky, M.~ElKawkagy, and A.~Keshk, ``Homomorphic encryption the “holy grail” of cryptography,'' in \emph{2016 2nd IEEE International Conference on Computer and Communications (ICCC)}, 2016, pp. 196--201, accessed: 2024-03-06.

\bibitem{pulido2021privacy}
B.~Pulido-Gaytan, A.~Tchernykh, J.~M. Cort{\'e}s-Mendoza, M.~Babenko, G.~Radchenko, A.~Avetisyan, and A.~Y. Drozdov, ``Privacy-preserving neural networks with homomorphic encryption: C hallenges and opportunities,'' \emph{Peer-to-Peer Networking and Applications}, vol.~14, no.~3, pp. 1666--1691, 2021.

\bibitem{njungle2025guardianml}
N.~B. Njungle, E.~Jahns, Z.~Wu, L.~Mastromauro, M.~Stojkov, and M.~Kinsy, ``Guardianml: Anatomy of privacy-preserving machine learning techniques and frameworks,'' \emph{IEEE Access}, 2025.

\bibitem{banerjee2019empirical}
C.~Banerjee, T.~Mukherjee, and E.~Pasiliao~Jr, ``An empirical study on generalizations of the relu activation function,'' in \emph{Proceedings of the 2019 ACM Southeast Conference}, 2019, pp. 164--167.

\bibitem{sharma2017activation}
S.~Sharma, S.~Sharma, and A.~Athaiya, ``Activation functions in neural networks,'' \emph{Towards Data Sci}, vol.~6, no.~12, pp. 310--316, 2017.

\bibitem{Podschwadt}
R.~Podschwadt, D.~Takabi, P.~Hu, M.~H. Rafiei, and Z.~Cai, ``A survey of deep learning architectures for privacy-preserving machine learning with fully homomorphic encryption,'' \emph{IEEE Access}, vol.~10, pp. 117\,477--117\,500, 2022.

\bibitem{catalano2005multiparty}
D.~Catalano, R.~Cramer, G.~Di~Crescenzo, I.~Darmg{\aa}rd, D.~Pointcheval, T.~Takagi, R.~Cramer, and I.~Damg{\aa}rd, ``Multiparty computation, an introduction,'' \emph{Contemporary cryptology}, pp. 41--87, 2005.

\bibitem{gazelle}
\BIBentryALTinterwordspacing
C.~Juvekar, V.~Vaikuntanathan, and A.~Chandrakasan, ``{GAZELLE}: A low latency framework for secure neural network inference,'' in \emph{27th USENIX Security Symposium (USENIX Security 18)}.\hskip 1em plus 0.5em minus 0.4em\relax Baltimore, MD: USENIX Association, Aug. 2018, pp. 1651--1669. [Online]. Available: \url{https://www.usenix.org/conference/usenixsecurity18/presentation/juvekar}
\BIBentrySTDinterwordspacing

\bibitem{liu2017oblivious}
J.~Liu, M.~Juuti, Y.~Lu, and N.~Asokan, ``Oblivious neural network predictions via minionn transformations,'' in \emph{Proceedings of the 2017 ACM SIGSAC conference on computer and communications security}, 2017, pp. 619--631.

\bibitem{riazi2019xonn}
M.~S. Riazi, M.~Samragh, H.~Chen, K.~Laine, K.~Lauter, and F.~Koushanfar, ``$\{$XONN$\}$:$\{$XNOR-based$\}$ oblivious deep neural network inference,'' in \emph{28th USENIX Security Symposium (USENIX Security 19)}, 2019, pp. 1501--1518.

\bibitem{gilad2016cryptonets}
R.~Gilad-Bachrach, N.~Dowlin, K.~Laine, K.~Lauter, M.~Naehrig, and J.~Wernsing, ``Cryptonets: Applying neural networks to encrypted data with high throughput and accuracy,'' in \emph{International conference on machine learning}.\hskip 1em plus 0.5em minus 0.4em\relax PMLR, 2016, pp. 201--210.

\bibitem{hesamifard2017cryptodl}
E.~Hesamifard, H.~Takabi, and M.~Ghasemi, ``Cryptodl: Deep neural networks over encrypted data,'' \emph{arXiv preprint arXiv:1711.05189}, 2017.

\bibitem{al2020towards}
A.~Al~Badawi, C.~Jin, J.~Lin, C.~F. Mun, S.~J. Jie, B.~H.~M. Tan, X.~Nan, K.~M.~M. Aung, and V.~R. Chandrasekhar, ``Towards the alexnet moment for homomorphic encryption: Hcnn, the first homomorphic cnn on encrypted data with gpus,'' \emph{IEEE Transactions on Emerging Topics in Computing}, vol.~9, no.~3, pp. 1330--1343, 2020.

\bibitem{e2dm}
\BIBentryALTinterwordspacing
X.~Jiang, M.~Kim, K.~Lauter, and Y.~Song, ``Secure outsourced matrix computation and application to neural networks,'' Cryptology {ePrint} Archive, Paper 2018/1041, 2018. [Online]. Available: \url{https://eprint.iacr.org/2018/1041}
\BIBentrySTDinterwordspacing

\bibitem{DiNN}
\BIBentryALTinterwordspacing
F.~Bourse, M.~Minelli, M.~Minihold, and P.~Paillier, ``Fast homomorphic evaluation of deep discretized neural networks,'' Cryptology {ePrint} Archive, Paper 2017/1114, 2017. [Online]. Available: \url{https://eprint.iacr.org/2017/1114}
\BIBentrySTDinterwordspacing

\bibitem{tfhe}
\BIBentryALTinterwordspacing
I.~Chillotti, N.~Gama, M.~Georgieva, and M.~Izabachène, ``Tfhe: Fast fully homomorphic encryption over the torus,'' Cryptology ePrint Archive, Paper 2018/421, 2018, \url{https://eprint.iacr.org/2018/421}. [Online]. Available: \url{https://eprint.iacr.org/2018/421}
\BIBentrySTDinterwordspacing

\bibitem{badawi2023}
\BIBentryALTinterwordspacing
A.~A. Badawi and Y.~Polyakov, ``Demystifying bootstrapping in fully homomorphic encryption,'' Cryptology {ePrint} Archive, Paper 2023/149, 2023. [Online]. Available: \url{https://eprint.iacr.org/2023/149}
\BIBentrySTDinterwordspacing

\bibitem{narumanchi2017performance}
H.~Narumanchi, D.~Goyal, N.~Emmadi, and P.~Gauravaram, ``Performance analysis of sorting of fhe data: integer-wise comparison vs bit-wise comparison,'' in \emph{2017 IEEE 31st International Conference on Advanced Information Networking and Applications (AINA)}.\hskip 1em plus 0.5em minus 0.4em\relax IEEE, 2017, pp. 902--908.

\bibitem{ckks}
\BIBentryALTinterwordspacing
J.~H. Cheon, A.~Kim, M.~Kim, and Y.~Song, ``Homomorphic encryption for arithmetic of approximate numbers,'' Cryptology {ePrint} Archive, Paper 2016/421, 2016. [Online]. Available: \url{https://eprint.iacr.org/2016/421}
\BIBentrySTDinterwordspacing

\bibitem{lee2022privacy}
J.-W. Lee, H.~Kang, Y.~Lee, W.~Choi, J.~Eom, M.~Deryabin, E.~Lee, J.~Lee, D.~Yoo, Y.-S. Kim, \emph{et~al.}, ``Privacy-preserving machine learning with fully homomorphic encryption for deep neural network,'' \emph{iEEE Access}, vol.~10, pp. 30\,039--30\,054, 2022.

\bibitem{kim2023optimized}
D.~Kim and C.~Guyot, ``Optimized privacy-preserving cnn inference with fully homomorphic encryption,'' \emph{IEEE Transactions on Information Forensics and Security}, vol.~18, pp. 2175--2187, 2023.

\bibitem{rovida_cnn}
\BIBentryALTinterwordspacing
L.~Rovida and A.~Leporati, ``Encrypted image classification with low memory footprint using fully homomorphic encryption,'' Cryptology {ePrint} Archive, Paper 2024/460, 2024. [Online]. Available: \url{https://eprint.iacr.org/2024/460}
\BIBentrySTDinterwordspacing

\bibitem{he_init}
E.~F. Brickell and Y.~Yacobi, ``On privacy homomorphisms (extended abstract),'' in \emph{Advances in Cryptology --- EUROCRYPT' 87}, D.~Chaum and W.~L. Price, Eds.\hskip 1em plus 0.5em minus 0.4em\relax Berlin, Heidelberg: Springer Berlin Heidelberg, 1988, pp. 117--125.

\bibitem{gentry}
C.~Gentry, ``A fully homomorphic encryption scheme,'' Ph.D. dissertation, Stanford University, 2009, \url{crypto.stanford.edu/craig}.

\bibitem{bgv}
N.~Aggarwal, C.~Gupta, and I.~Sharma, ``Fully homomorphic symmetric scheme without bootstrapping,'' pp. 14--17, 2014.

\bibitem{bfv}
\BIBentryALTinterwordspacing
J.~Fan and F.~Vercauteren, ``Somewhat practical fully homomorphic encryption,'' \emph{IACR Cryptol. ePrint Arch.}, vol. 2012, p. 144, 2012. [Online]. Available: \url{https://api.semanticscholar.org/CorpusID:1467571}
\BIBentrySTDinterwordspacing

\bibitem{lwepropsed}
O.~Regev, ``On lattices, learning with errors, random linear codes, and cryptography,'' \emph{Procedings of the thirty-seventh annual ACM symposium on Theory of Computing}, 2005.

\bibitem{ringlwe}
\BIBentryALTinterwordspacing
V.~Lyubashevsky, C.~Peikert, and O.~Regev, ``On ideal lattices and learning with errors over rings,'' Cryptology {ePrint} Archive, Paper 2012/230, 2012. [Online]. Available: \url{https://eprint.iacr.org/2012/230}
\BIBentrySTDinterwordspacing

\bibitem{MohammadAlRubaie-2019}
M.~Al-Rubaie and J.~M. Chang, ``Privacy-preserving machine learning: Threats and solutions,'' \emph{IEEE Security \& Privacy}, vol.~17, no.~2, pp. 49--58, 2019.

\bibitem{ZoubinGhahramani-2004}
Z.~Ghahramani, \emph{Unsupervised Learning}.\hskip 1em plus 0.5em minus 0.4em\relax Berlin, Heidelberg: Springer Berlin Heidelberg, 2004, pp. 72--112.

\bibitem{TammyJiang-2020}
T.~Jiang, J.~L. Gradus, and A.~J. Rosellini, ``\BIBforeignlanguage{en}{Supervised machine learning: A brief primer},'' \emph{\BIBforeignlanguage{en}{Behav Ther}}, vol.~51, no.~5, pp. 675--687, May 2020.

\bibitem{KeerthanaSivamayil-2023}
\BIBentryALTinterwordspacing
K.~Sivamayil, E.~Rajasekar, B.~Aljafari, S.~Nikolovski, S.~Vairavasundaram, and I.~Vairavasundaram, ``A systematic study on reinforcement learning based applications,'' \emph{Energies}, vol.~16, no.~3, 2023. [Online]. Available: \url{https://www.mdpi.com/1996-1073/16/3/1512}
\BIBentrySTDinterwordspacing

\bibitem{AlexKrizhevsky-2012}
\BIBentryALTinterwordspacing
A.~Krizhevsky, I.~Sutskever, and G.~E. Hinton, ``Imagenet classification with deep convolutional neural networks,'' in \emph{Advances in Neural Information Processing Systems}, F.~Pereira, C.~Burges, L.~Bottou, and K.~Weinberger, Eds., vol.~25.\hskip 1em plus 0.5em minus 0.4em\relax Curran Associates, Inc., 2012. [Online]. Available: \url{https://papers.nips.cc/paper\_files/paper/2012/ file/c399862d3b9d6b76c8436e924a68c45b-Paper.pdf}
\BIBentrySTDinterwordspacing

\bibitem{DUBEY202292}
\BIBentryALTinterwordspacing
S.~R. Dubey, S.~K. Singh, and B.~B. Chaudhuri, ``Activation functions in deep learning: A comprehensive survey and benchmark,'' \emph{Neurocomputing}, vol. 503, pp. 92--108, 2022. [Online]. Available: \url{https://www.sciencedirect.com/science/article/pii/S0925231222008426}
\BIBentrySTDinterwordspacing

\bibitem{roodschild2020new}
M.~Roodschild, J.~Gotay~Sardi{\~n}as, and A.~Will, ``A new approach for the vanishing gradient problem on sigmoid activation,'' \emph{Progress in Artificial Intelligence}, vol.~9, no.~4, pp. 351--360, 2020.

\bibitem{hu2018overcoming}
Y.~Hu, A.~Huber, J.~Anumula, and S.-C. Liu, ``Overcoming the vanishing gradient problem in plain recurrent networks,'' \emph{arXiv preprint arXiv:1801.06105}, 2018.

\bibitem{gupta2020effect}
N.~Gupta, P.~Bedi, and V.~Jindal, ``Effect of activation functions on the performance of deep learning algorithms for network intrusion detection systems,'' in \emph{Proceedings of ICETIT 2019: Emerging Trends in Information Technology}.\hskip 1em plus 0.5em minus 0.4em\relax Springer, 2020, pp. 949--960.

\bibitem{he2018relu}
J.~He, L.~Li, J.~Xu, and C.~Zheng, ``Relu deep neural networks and linear finite elements,'' \emph{arXiv preprint arXiv:1807.03973}, 2018.

\bibitem{gil2012non}
L.~A. Gil-Alana and J.~C. Cuestas, ``A non-linear approach with long range dependence based on chebyshev polynomials,'' 2012.

\bibitem{fhew}
\BIBentryALTinterwordspacing
I.~Chillotti, N.~Gama, M.~Georgieva, and M.~Izabachène, ``Faster fully homomorphic encryption: Bootstrapping in less than 0.1 seconds,'' Cryptology {ePrint} Archive, Paper 2016/870, 2016. [Online]. Available: \url{https://eprint.iacr.org/2016/870}
\BIBentrySTDinterwordspacing

\bibitem{chimera}
\BIBentryALTinterwordspacing
C.~Boura, N.~Gama, M.~Georgieva, and D.~Jetchev, ``{CHIMERA}: Combining ring-{LWE}-based fully homomorphic encryption schemes,'' Cryptology {ePrint} Archive, Paper 2018/758, 2018. [Online]. Available: \url{https://eprint.iacr.org/2018/758}
\BIBentrySTDinterwordspacing

\bibitem{OpenFHE}
\BIBentryALTinterwordspacing
A.~A. Badawi, J.~Bates, F.~Bergamaschi, D.~B. Cousins, S.~Erabelli, N.~Genise, S.~Halevi, H.~Hunt, A.~Kim, Y.~Lee, Z.~Liu, D.~Micciancio, I.~Quah, Y.~Polyakov, S.~R.V., K.~Rohloff, J.~Saylor, D.~Suponitsky, M.~Triplett, V.~Vaikuntanathan, and V.~Zucca, ``Openfhe: Open-source fully homomorphic encryption library,'' Cryptology ePrint Archive, Paper 2022/915, 2022, \url{https://eprint.iacr.org/2022/915}. [Online]. Available: \url{https://eprint.iacr.org/2022/915}
\BIBentrySTDinterwordspacing

\bibitem{openfhesec}
\BIBentryALTinterwordspacing
A.~Al~Badawi, J.~Bates, F.~Bergamaschi, D.~B. Cousins, S.~Erabelli, N.~Genise, S.~Halevi, H.~Hunt, A.~Kim, Y.~Lee, Z.~Liu, D.~Micciancio, I.~Quah, Y.~Polyakov, S.~R.V., K.~Rohloff, J.~Saylor, D.~Suponitsky, M.~Triplett, V.~Vaikuntanathan, and V.~Zucca, ``Openfhe: Open-source fully homomorphic encryption library,'' in \emph{Proceedings of the 10th Workshop on Encrypted Computing \& Applied Homomorphic Cryptography}, ser. WAHC'22.\hskip 1em plus 0.5em minus 0.4em\relax New York, NY, USA: Association for Computing Machinery, 2022, pp. 53--63. [Online]. Available: \url{https://doi.org/10.1145/3560827.3563379}
\BIBentrySTDinterwordspacing

\bibitem{lenets}
Y.~Lecun, L.~Bottou, Y.~Bengio, and P.~Haffner, ``Gradient-based learning applied to document recognition,'' \emph{Proceedings of the IEEE}, vol.~86, no.~11, pp. 2278--2324, 1998.

\bibitem{he2015deepresiduallearningimage}
\BIBentryALTinterwordspacing
K.~He, X.~Zhang, S.~Ren, and J.~Sun, ``Deep residual learning for image recognition,'' 2015. [Online]. Available: \url{https://arxiv.org/abs/1512.03385}
\BIBentrySTDinterwordspacing

\bibitem{solsupercom}
\BIBentryALTinterwordspacing
D.~M. Jennewein, J.~Lee, C.~Kurtz, W.~Dizon, I.~Shaeffer, A.~Chapman, A.~Chiquete, J.~Burks, A.~Carlson, N.~Mason, A.~Kobawala, T.~Jagadeesan, P.~B. Basani, T.~Battelle, R.~Belshe, D.~McCaffrey, M.~Brazil, C.~Inumella, K.~Kuznia, J.~Buzinski, D.~D. Shah, S.~M. Dudley, G.~Speyer, and J.~Yalim, ``The sol supercomputer at arizona state university,'' ser. PEARC '23.\hskip 1em plus 0.5em minus 0.4em\relax New York, NY, USA: Association for Computing Machinery, 2023, p. 296–301. [Online]. Available: \url{https://doi.org/10.1145/3569951.3597573}
\BIBentrySTDinterwordspacing

\end{thebibliography}
\end{document}